# Effect of layer thickness on the mechanical behavior of oxidation-strengthened Zr/Nb nanoscale multilayers

M. A. Monclús[1,*], M. Callisti[2,*,#], T. Polcar[3,4], L. W. Yang[1,5], J. M. Molina-Aldareguía[1,*] and J. LLorca[1,5]

[1] IMDEA Materials Institute, c/Eric Kandel 2, 28906 Getafe, Madrid, Spain.
[2] Engineering Materials, Faculty of Engineering and the Environment, University of Southampton, Southampton SO17 1BJ, UK.
[3] National Centre for Advanced Tribology (nCATS), Faculty of Engineering and the Environment, University of Southampton, Southampton SO17 1BJ, UK
[4] Department of Control Engineering, Faculty of Electrical Engineering, Czech Technical University in Prague, Technická 2, Prague 6, Czech Republic.
[5] Department of Materials Science, Universidad Politécnica de Madrid. E.T.S. de Ingenieros de Caminos, 28040 Madrid, Spain.

**Abstract**

The effect of bilayer thickness ($L$) reduction on the oxidation-induced strengthening of Zr/Nb nanoscale metallic multilayers (NMM) is investigated. Zr/Nb NMMs with L = 10 and 75 nm were annealed at 350 ºC for a time ranging between 2 and 336 h and the changes in structure and deformation behaviour were studied by nano-scale mechanical testing and analytical electron microscopy. Annealing led to transformation of the Zr layers into $ZrO_2$ after a few hours, while the Nb layers oxidised progressively at a much slower rate. The sequential oxidation of Zr and Nb layers was found to be key for the oxidation to take place without rupture of the multi-layered structure and without coating spallation in all cases. However, the multilayers with the smallest bilayer thickness ($L$=10 nm) presented superior damage tolerance and therefore structural integrity during the oxidation process, while for $L$=75 nm the volumetric expansion associated with oxidation led to the formation of cracks at the interfaces and within the $ZrO_2$ layers. As a result, the nanoindentation hardness increase after annealing was significantly higher for the nanolaminate with $L$ = 10 nm. Comparison between nanoindentation and micropillar compression behaviour of the oxidized NMMs demonstrate that the hardness increase upon oxidation arises from the contribution of the residual stresses associated with the volume increase due to oxidation and to the higher strength of the oxides.

* Corresponding authors
# Current address: Department of Materials Science and Metallurgy, Cambridge University, 27 Charles Babbage Road, Cambridge, CB3 0FS, United Kingdom

# 1. Introduction

Nanoscale metallic multilayers (NMM) with individual layer thickness below ≈ 100 nm have been widely studied due to their outstanding mechanical properties, that arise from the high density of interfaces and the nanoscale dimensions of the layers [1,2]. There are many possible material combination as a function of crystal structure and preferred texture of each layer that result in different interface types [3] (i.e. coherent, semi-coherent and incoherent interfaces) and thus NMMs stand for perfect candidates to analyze the effect of interfaces and crystallographic orientation on the mechanisms of plastic deformation at the nm scale.

Moreover, NMMs are promising material systems for future applications in various fields ranging from electronics and computer science (micro/nano-electromechanical systems [2], magnetic data storage [4, 5]), hydrogen storage [6, 7], tribology [8] and nuclear industry [9-16]. Most of these applications require properties such as high strength, toughness, wear resistance and structural stability in demanding environments (high temperature, harsh atmosphere, etc.). However, in spite of the large number of studies focusing on the correlation between structure (in as-produced condition) and mechanical properties of NMMs [17-33], very little information is available on the thermal stability and the high temperature properties of NMMs, especially under oxidising environments (not vacuum). In particular, among the cubic systems, the extensively studied Cu/Nb multilayers, which possess high strength [34–36], radiation tolerance and good thermal stability [37, 38] when annealed in vacuum or in inert atmospheres, are prone to degradation in air at temperatures as low as 300°C [39]. Other systems, like Cu/W, Cu/Cr and Cu/Mo exhibit a dramatic drop in hot hardness for T > 200°C due to stress assisted diffusional flow of Cu [20,21] and a gradual degradation of the layered architecture and formation of a pronounced globular structure after annealing at T ≥ 700°C in vacuum [40]. A similar result was observed on Ag/Ni NMMs [41] annealed in vacuum at a much lower temperature (300°C). On the other hand, Cu/Ta NMMs [42] retained the layered architecture at temperatures as high as 800°C, although a mixed boundary layer formed at interfaces, and no information was reported about the evolution of mechanical properties. Similar results have been reported for hcp-based systems, like Co/Cu NMMs [43] or Ni/Ru NMMs [44], which exhibited some structural degradation (i.e. Cu layers breakage) after annealing in vacuum at 400°C and 600°C, respectively.

Zr/Nb, a highly immiscible metal-metal hcp/bcc combination [45], constitutes, however, a remarkable system from the point of view of high temperature behaviour under oxidizing environments. Contrary to other NMMs, it was recently shown that, for a Zr/Nb multilayer with layer thickness (45 nm/30 nm), long term annealing at 300 – 400 ºC in air resulted in the oxidation of the Zr and Nb layers, in such a way that a layered structure with sharp interfaces was preserved, without coating spallation [46]. This was possible because oxidation of the Zr and Nb layers took place sequentially. As shown in [46], a rapid oxidation of the Zr layers to $ZrO_2$ was observed, and the large compressive residual stresses generated as a result of the 56 % volumetric expansion during oxidation of Zr, were partially relieved by the plastic deformation of the metallic Nb layers, that at this point were left under tensile residual stresses. Oxidation of the Nb layers to $Nb_2O_5$ was, however, slower but implied a 154% volumetric expansion, reducing the compressive stresses in the $ZrO_2$ layers and introducing high compressive stresses in the $Nb_2O_5$ layers. As a result, the oxidation process led to the development of an oxide/metal nanocomposite coating with high strength and good thermal stability. This observation makes Zr-based NMMs promising candidate materials for the future nuclear industry [47], since the interfaces are known to play a positive role against radiation damage [10,48]. However, the role of individual layer thickness on the deformation mechanisms of oxidised multilayers as a function of the corresponding structural changes has not yet been reported. Therefore, in this study the effect of oxidation (after annealing from 2 to 336 h at 350 ºC) on the mechanical behavior of Zr/Nb NMMs was studied as a function of the bilayer thickness ($L$ = 10 and 75 nm). Nanoindentation and micropillar compression tests, together with high-temperature X-ray diffraction (HT-XRD) and analytical transmission electron microscopy (ATEM), were used to shed light on deformation and failure mechanisms in Zr/Nb NMMs before and after oxidation.

## 2. Materials and experimental methods

*2.1. Materials synthesis*

Zr/Nb multilayers were deposited on single crystal (100) Si wafers using a balanced magnetron sputtering apparatus (Kurt J. Lesker Company, Pennsylvania, US) with bilayer layer thicknesses ($L$) of 10 and 75 nm and total layer thicknesses of ≈ 1.35 μm and ≈ 1.1 μm respectively. Monolithic Zr and Nb films were also deposited on the same substrate with thicknesses of 0.85 μm and 1.24 μm, respectively. Further details about the deposition process are reported elsewhere [49]. The multilayer and monolithic films were thermally annealed in air inside a muffle furnace at 350°C for annealing times $t_a$ = 2, 15, 48, 168 and 336 hrs. The as-deposited and annealed Zr/Nb multilayers presented a roughness, $R_a$, measured by AFM, between 3 - 5 nm.

*2.2. Structural and chemical characterisation*

High-temperature X-ray diffraction (HT-XRD) analyses were carried out by using a Philips diffractometer (Co Kα-radiation) in Bragg-Brentano geometry. Samples (10 × 10 mm) were fixed on a Pt plate and heated from room temperature up to 400°C in steps of 100°C. The heating rate was 40°C/min between each target temperature at which every sample was kept for 1 h before XRD data acquisition in the 2θ range 20–70°. XRD measurements were performed in a vacuum chamber, specifically designed for *in situ* studies, coupled with the diffractometer. In order to study the progressive oxidation process (avoiding a fast oxidation of the samples), the overall oxidation rate of Zr/Nb samples was decreased by performing high temperature XRD analyses in a protective atmosphere. To this aim, experiments were performed under continuous gas (Ar–5%$H_2$) flux at low pressure (1–10 Pa) after evacuating the chamber down to a pressure lower than $10^{-3}$ Pa. Finer structural analyses on as-deposited and annealed multilayers as well as on compressed pillars were carried out by using a CS probe-corrected JEOL ARM200F (cold-FEG) TEM/STEM operated at 200 kV and equipped with a 100 $mm^2$ Centurion EDX detector (Thermo Fisher Scientific Inc., Madison, Wisconsin, USA). TEM samples were prepared by using a focused ion beam (FIB) system (FEI Helios 600i).

*2.3. Mechanical characterisation*

Nanoindentation measurements were performed using the Nanotest Platform 3 instrument (Micromaterials, Wrexham, UK) equipped with a diamond Berkovich indenter. The hardness values reported below are an average of ten indents performed by using a maximum load of 5 mN (for the Zr/Nb multilayer) and 2 mN (for monolithic Zr and Nb coatings and high temperature tests) with loading, holding and unloading times of 20, 10 and 5 s, respectively. The maximum load was chosen in such a way that $h_c/t < 0.1$, where $h_c$ is the contact depth and *t* the film thickness. High temperature nanoindentation tests were also performed at T = 100, 200, 300 and 350°C; the indentation chamber was filled with Ar gas until a final $O_2$ level of < 0.2 % was reached. The sample was mounted on a hot stage using high temperature cement (Omegabond 600). A thermocouple was placed on the sample surface and used to match the temperature between the sample surface and the indenter tip in order to minimize thermal drift.

Square micropillars of the Zr/Nb NMs (*L* = 10 and 75 nm) were machined in as-deposited coatings as well as in coatings annealed for 48 hrs at 350ºC, using a FEI Helios Nanolab 600i dualbeam FIB-FEGSEM operated at 30 kV. In the first stage, a large square of 30 μm outer side and 10 μm inner side was milled at currents of 21 nA to provide sufficient clearance for the flat-punch indenter. A series of lower currents milling steps were performed in the second stage until a final fine milling step with a current of 48 pA, to minimise the damage caused by the Ga$^+$ ions. The length of the side of the square micropillar was ≈ 650–750 nm, and the aspect ratio (length/side) was ≈ 2.0–2.2 while the taper of the pillars was between 3 and 5°. The taper was further reduced by tilting the specimen an additional 1–2° with respect to the 52° milling position using a beam current of 24 pA, resulting in a final pillar taper below 1°.

Micropillar compression tests were performed with the Hysitron TI950 TriboIndenter equipped with a 10 μm-diameter flat punch diamond indenter. All micropillars were compressed under displacement control at loading rates of 1.2 and 12 nm/s (corresponding to constant strain rates of ≈ 1×10$^{-3}$–1×10$^{-2}$ s$^{-1}$) and strains between 10 and 50 %. The displacement of the indenter tip includes the compression of the base of the pillar into the substrate, and Sneddon´s correction (which considers that the pillar acts as a perfectly rigid flat punch indenter pushing into an isotropic half space [50]) was used to compute the micropillar deformation. The pillars were considered taper-free with cross-sectional area ($A_0$) equal to the top surface of the pillars and length $L_0$. The load-displacement curves were converted to engineering stress ($\sigma_{eng}$) - engineering strain ($\varepsilon_{eng}$) curves according to

$$\sigma_{eng} = \frac{P}{A_0} \tag{1}$$

$$\varepsilon_{eng} = \frac{d_p}{L_0} \tag{2}$$

where $P$ is the applied force and $d_p$ is the pillar displacement. Images of the deformed pillars were obtained using the FEI Helios Nanolab 600i dual beam FIB-FEGSEM.

## 3. Results and discussion

### 3.1. Microstructure of Zr/Nb multilayers

Detailed structural characterisation of Zr/Nb films ($L$ = 10 and 75 nm) was performed by *in-situ* XRD in order to shed light on the oxidation process of Zr/Nb NMMs as a function of the temperature in the range 25°C–400°C. In addition, the effect of annealing time (2, 48 and 168 hrs) at 350°C was investigated by means of analytical transmission electron microscopy. The microstructure of compressed micropillars produced from as-deposited and annealed films was also investigated and discussed in detail.

### 3.1.1. In situ HT-XRD analyses

XRD analyses of Zr/Nb NMMs ($L$ = 10 and 75 nm) acquired *in situ* at different temperatures (25ºC–400°C) are shown in Fig. 1. The XRD patterns in Fig. 1a ($L$ = 10 nm) are densely populated by phase and satellite peaks due to the superlattice structure of the multilayer. These satellites around the main diffraction peaks (i.e. $(0002)_{Zr}$ and $(110)_{Nb}$) became less intense as the temperature increased, and this result implies a degradation of the superlattice structure and changes at the interfaces. Many overlapping peaks populate the XRD pattern at 300ºC and they are clearer at 400°C. These peaks were attributed to $ZrO_2$ (monoclinic) and $Nb_2O_5$ (monoclinic) phases and this is consistent with findings reported in our previous study [46], where EELS quantitative analyses suggested formation of $ZrO_2$ and $Nb_2O_5$ compounds. The main phases for Zr and Nb were still observed after heating at 400°C, which suggests a partial oxidation of the multilayer, due to the low oxidation rates encountered at the low oxygen partial pressure at which the *in-situ* tests were performed.

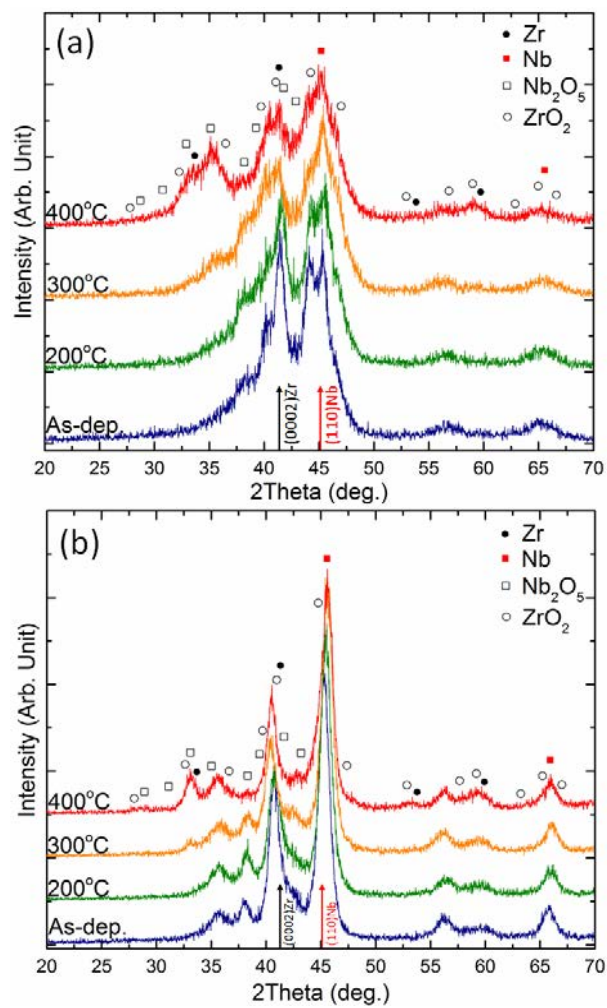

**Fig. 1.** In situ HT-XRD patterns of Zr/Nb NMMs with a bilayer periodicity of (a) 10 nm and (b) 75 nm annealed in vacuum at different temperatures up to 400°C.

The Zr/Nb NMM with $L$ = 75 nm also showed the formation of the same oxides mentioned above (Fig. 1b). Based on a qualitative analysis of the patterns, weak $ZrO_2$ peaks appeared already at 200°C, while some $Nb_2O_5$ peaks were noticed unambiguously at 300°C. This observation is likely to be related to the fact that Zr is the top layer in these NMMs, while Nb was found to exhibit a much slower oxidation rate [46]. A noticeable and opposite peak shift was observed for $(0002)_{Zr}$ and $(110)_{Nb}$ suggesting a concomitant compression of the $(0002)_{Zr}$ planes and expansion of $(110)_{Nb}$ planes. Temperatures between 300–400°C were found to be critical for initiation and development of the oxidation process in both NMMs.

3.1.2. Analytical electron microscopy

HAADF-STEM images of complete cross-sections and magnified views of the region between the severely oxidised and the lightly or non-oxidised regions (hereafter defined as boundary) are shown in Fig. 2 for Zr/Nb NMMs with $L$ = 10 nm annealed for 2, 48 and 168 hrs at 350ºC. Based on the observed HAADF contrast, the first 10 layers were affected by the oxidation process after annealing for 2 hrs, while the oxidised region extended to 25/26 and 35/36 layers for annealing times of 48 and 168 hrs, respectively. The number of oxidised layers was found to increase following a logarithmic law in relation to the annealing time, as expected for a diffusion controlled process. The oxidation process caused a thickness increase from ∼ 1.09 μm (as-deposited and after 2 hrs annealing) to 1.25 μm (after 48 hrs) and 1.3 μm (after 168 hrs). A blurred region was very often observed across the boundary, which was attributed to local disruption of the layered structure. An increasing number of layers were involved at the boundary as a function of the annealing time (see Figs. 2d-f). The observed disruption was caused by the high thermal stress mismatch developed at the boundary between the oxidised top region and the underlying region. Overall, the NMMs with $L$ = 10 nm showed a good structural integrity even after annealing for 168 hrs. In fact, no interfacial delamination was observed across the layers. Furthermore, no crack propagation of existing defects along columnar grain boundaries was triggered by the high residual stresses developed during annealing due to the volume increase associated with oxidation.

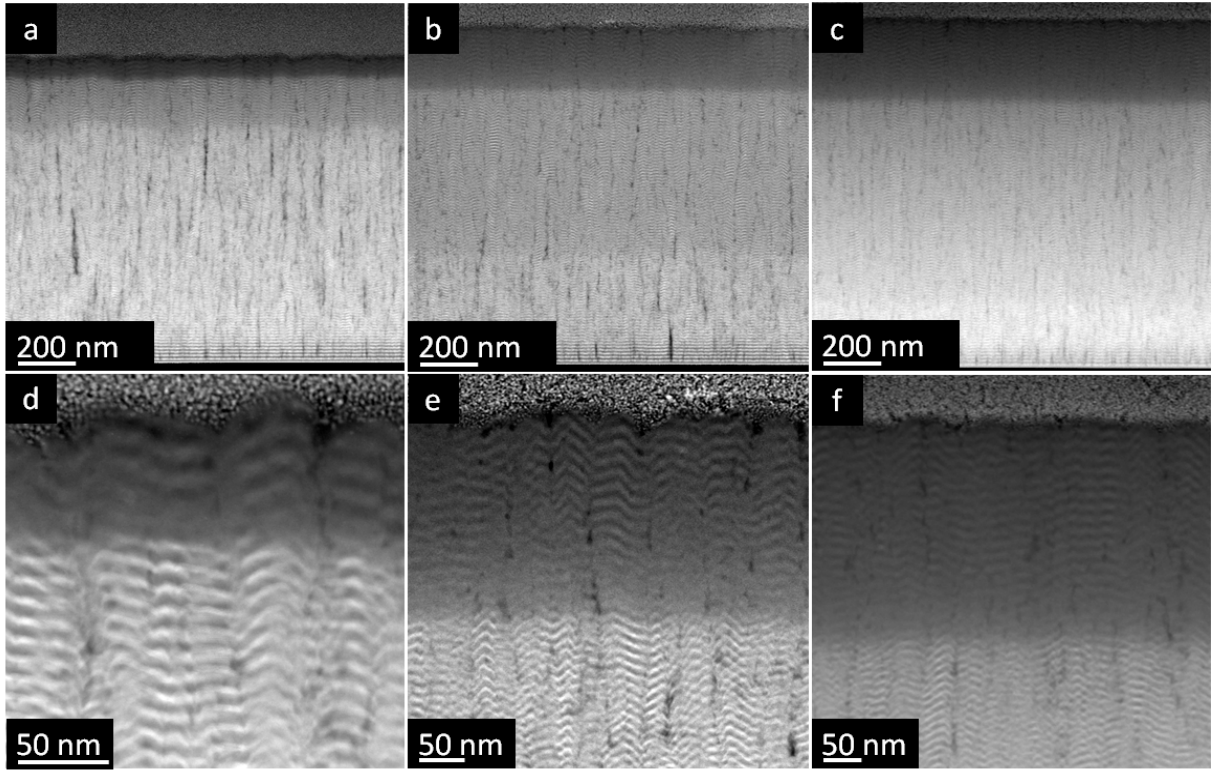

**Fig. 2.** Cross-sectional HAADF-STEM images of Zr/Nb NMMs ($L = 10$ nm) annealed at 350°C for: (a,d) 2 hrs, (b,e) 48 hrs and (c,f) 168 hrs. Figures (a, b and c) show the whole cross-section, while figures (d, e and f) show magnified views of the boundary between the heavily oxidised and non-oxidised regions.

A rather complex scenario was found at the atomic scale especially after annealing for longer times. The atomic structure and nature of interfaces well below the oxidation boundary did not change after annealing, i.e. $(110)_{Nb}//(0002)_{Zr}$ as reported in previous studies [45,46]. Therefore, the analyses in the current study were focused on the boundary region and in the oxidised regions where major changes were observed. The atomic arrangement of the constituent layers for different annealing times is shown in Fig. 3. Some Zr layers were transformed into $ZrO_2$ (monoclinic) after annealing for 2 hrs (Fig. 3a), while Nb layers transformed mostly into $Nb_2O_5$ (monoclinic). The fast Fourier transform (FFT) (inset in Fig. 3a) shows that $(\bar{1}11)_{ZrO_2}//(111)_{Nb_2O_5}$ (parallel to the substrate) in the oxidised region. Weak diffraction spots were also attributed to $(400)_{Nb_2O_5}$. After annealing for 48 hrs, $(\bar{1}11)_{ZrO_2}$ and $(111)_{Nb_2O_5}$ were still identified from the diffracting spots.

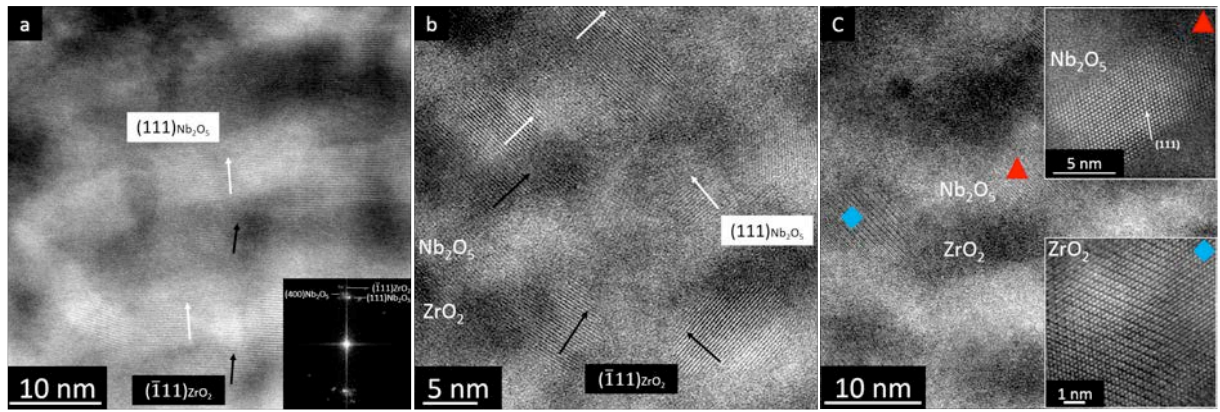

**Fig. 3.** High-resolution HAADF-STEM images of Zr/Nb NMMs ($L$ = 10 nm) annealed at 350°C for: (a) 2 hrs, (b) 48 hrs and (c) 168 hrs. In (b) arrows indicate the direction normal to $(111)Nb_2O_5$ and $(\bar{1}11)ZrO_2$ planes. In (c) triangles and rhombus indicate the regions where the images in the insets were acquired.

The strong texture observed in the film annealed for 2 hrs (see inset in Fig. 3a) was, however, weakened in the film annealed for 48 hrs. Arrows in Fig. 3b indicate that crystalline domains in the constituent layers exhibited a more random crystallographic orientation with respect to the interface. The atomic structure of the film annealed for 168 hrs is shown in Fig. 3c, where an even more complex scenario was observed. In fact, crystalline domains ($ZrO_2$ and $Nb_2O_5$ phases – see inset in Fig. 3c) with a random orientation as well as amorphous regions were observed. Fig. 4 shows that most of the oxygen was stored in the region above the boundary and, to a lower extent, in some Zr layers below the boundary. The oxygen content gradually decreased across the film thickness toward the substrate. For both heat treatments, the layered structure was mostly retained in the heavily oxidised region, although ruptures along columnar grain boundaries may have caused some interconnection between layers as shown in the top region in Fig. 4a.

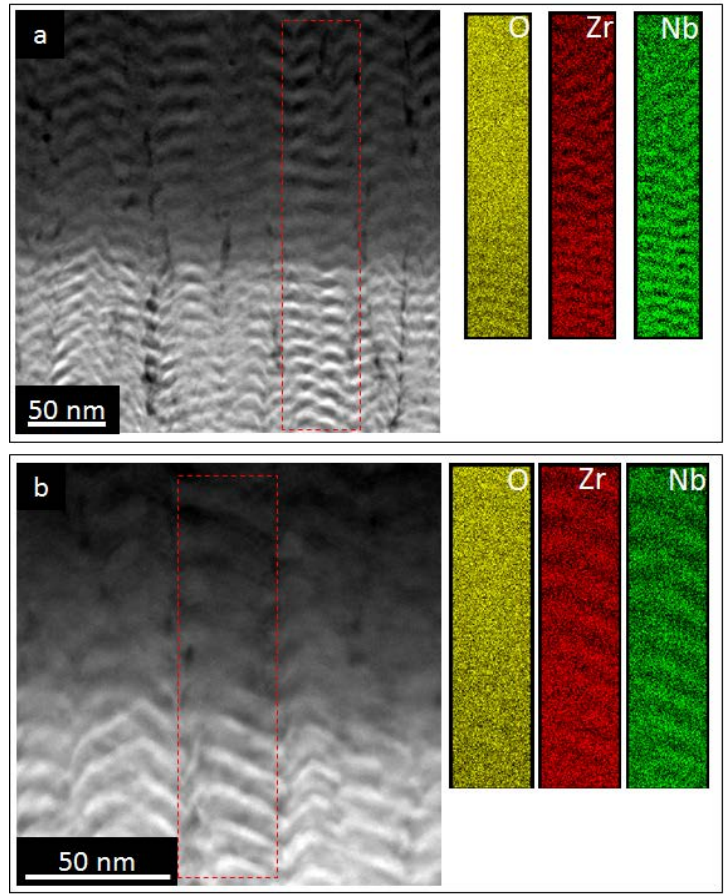

**Fig. 4.** Cross-sectional HAADF-STEM images and EDX maps of Zr/Nb NMMs ($L = 10$ nm) annealed at 350 °C for: (a) 48 hrs and (b) 168 hrs. Windows in the STEM images indicate the regions where EDX maps were acquired.

HAADF-STEM images of the cross-section of the Zr/Nb NMMs with $L = 75$ nm annealed for 2 and 168 hrs are shown in Fig. 5. An obvious increase in thickness was observed after annealing, from 1.32 μm (as-deposited) to 1.5 μm (for 2 hrs) and finally to 1.64 μm (for 168 hrs). In particular, annealing for 2 hrs caused the oxidation of 1.5 Nb layers where the oxidation boundary was located in the second Nb layer (Fig. 5c), while Zr layers were already fully oxidised near the top and partially oxidised towards the substrate. After 168 hrs, the number of Nb oxidised layers increased to 5 (again the boundary is located in a Nb layer – see Fig. 5d). Contrary to the NMM with $L = 10$ nm, for $L = 75$ nm, voids and pores inside the layers were found in the oxidised region (Fig. 5c) as well as cracks that initiated and propagated along interfaces (Fig. 5d). These defects were not prominent in the oxidized NMMs with $L = 10$ nm (Fig. 2).

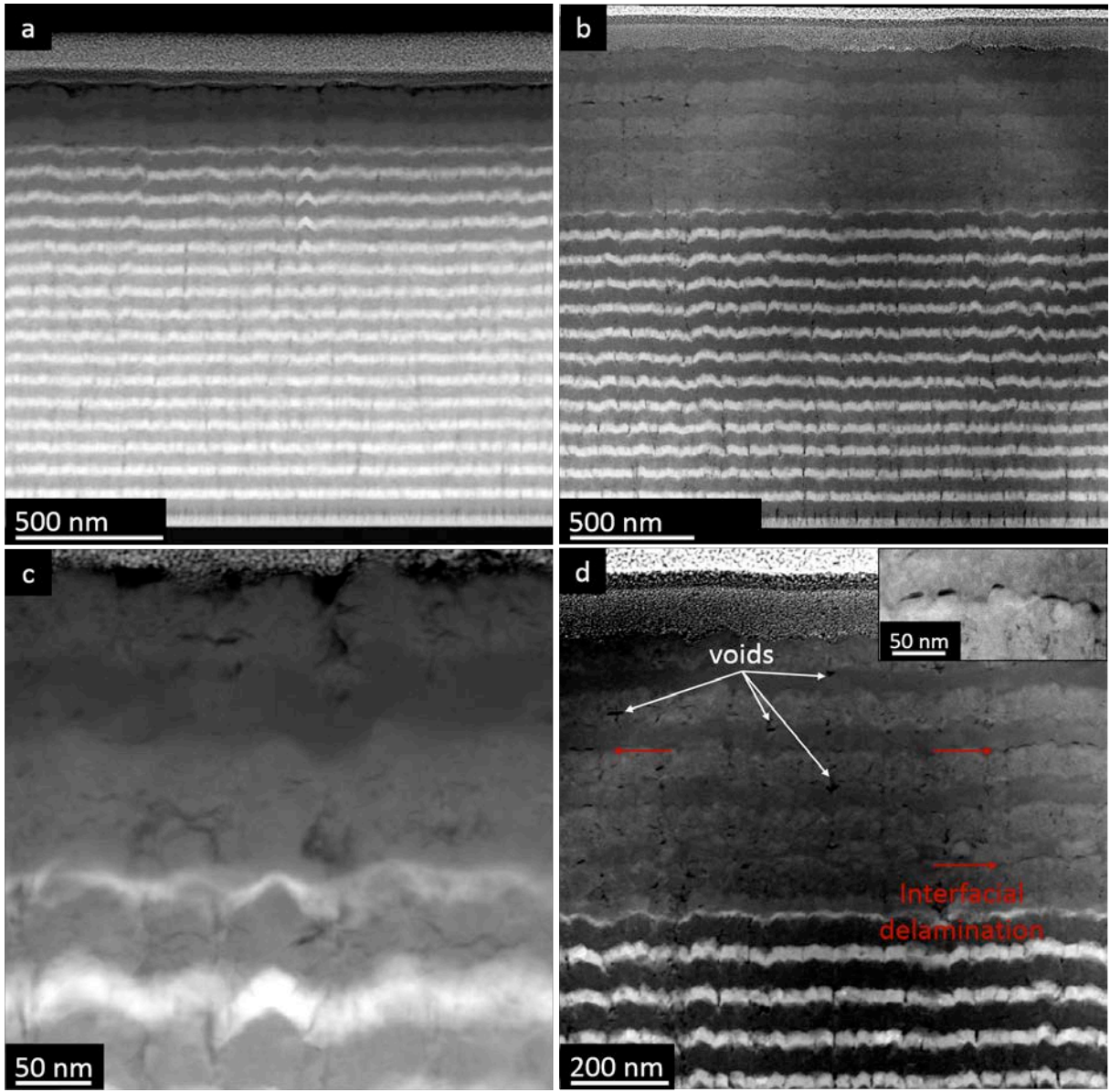

**Fig. 5.** Cross-sectional HAADF-STEM images of Zr/Nb NMMs ($L = 75$ nm) annealed at 350°C for: (a-c) 2 hrs, (b-d) 168 hrs. In (d) horizontal arrows (amaranth) point regions where interfacial delamination occurred (see inset in d), while the other arrows (white) points voids formed after annealing.

The oxygen distribution after annealing in the NMMs with $L = 75$ nm (Fig. 6) was very similar to the one found in Fig. 4 for NMMs with $L = 10$ nm, as well as the phase distribution, as can be found in more detail in [46]. In summary, the TEM results show that, irrespective of layer thickness, all the Zr layers were rapidly transformed into monoclinic $ZrO_2$, while the oxidation of Nb layers took much longer, and proceeded gradually from the top surface to the bottom.

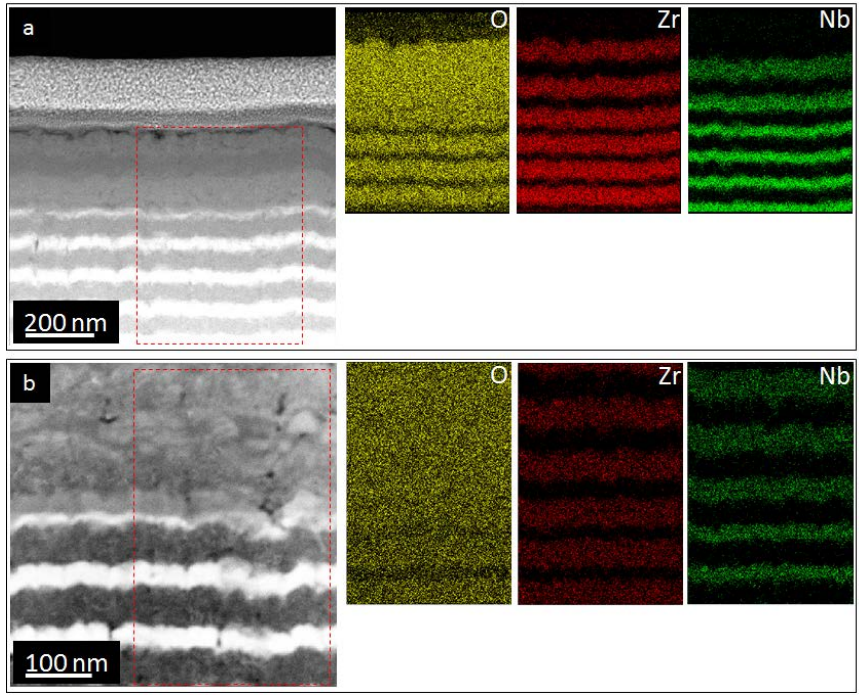

**Fig. 6.** Cross-sectional HAADF-STEM images and EDX maps of Zr/Nb NMMs ($L = 75$ nm) annealed at 350°C for: (a) 2 hrs and (b) 168 hrs. Windows in the STEM images indicate the regions where EDX maps were acquired.

*3.2. Mechanical properties*

3.2.1. Nanoindentation

The hardness of the Zr/Nb NMMs is plotted as a function of indentation temperature in Fig. 7a. As expected, hardness initially drops with temperature due to the reduction of the flow stress of the metallic Zr and Nb with temperature, as it was reported in other metallic NMMs [20]. The modulation periodicity had little effect on the room temperature hardness (indicated as "as deposited" in Fig. 7), as has been shown before and explained based on changes in the crystalline orientation of the Zr layers with layer thickness [45]. However, the hardness decreased more rapidly with temperature for $L = 10$ nm. This is a common observation in other metallic NMMs [20,39,51,52] and it is thought to arise due to the easier activation of stress assisted diffusional flow mechanisms for the thinnest multilayers, as a result of their higher grain boundary and interface densities. However, the hardness increased with temperature for T > 200°C for both NMMs. Although the experiments were carried out inside an Ar-purged chamber, the samples were kept at each temperature for more than an hour and the residual $O_2$ in the chamber led to the oxidation of the top layers, particularly for T > 200°C, as observed during in situ HT-XRD. After the high temperature testing, which lasted for about 6 hrs, the samples were allowed to cool down from T = 350°C to room temperature and the hardness for Zr/Nb with $L$ = 10 and 75 nm after annealing was ≈ 12.2 and 10 GPa, respectively (see "after cooling" values in Fig. 7a).

The hardness measured at room temperature after annealing of the Zr/Nb NMMs and monolithic Zr and Nb films as a function of annealing time ($t_a$) is shown in Fig. 7b. The hardness of Zr/Nb NMMs was ≈ 7.2 GPa for $L$ = 10 nm and ≈ 7.7 GPa for $L$ = 75 nm in the as-deposited condition. They are slightly larger than the one given by the rule of mixtures from the hardness of the Zr and Nb constituents ($H_{Zr}$ = 6.1 GPa and $H_{Nb}$ = 6.7 GPa) and this strengthening is typically found on metallic NMs with incoherent interfaces [1]. The actual deformation mechanisms of Zr/Nb NMs have been analysed previously [45,49] and it has been reported that the strength of the NMM with $L$ = 75 nm follows the confined layer slip (CLS) model whereas the strength is controlled by dislocation transmission across interfaces for $L$ = 10 nm.

The annealing treatment at 350°C of the Zr/Nb NMMs and the monolithic Zr and Nb films led to a large increase in hardness (Fig. 7b). The largest increase was found for Zr/Nb with $L$ = 10 nm and for the monolithic Zr layer: the hardness almost doubled after 2 hrs of annealing, following the progressive Zr oxidation. The monolithic Zr layer delaminated after a few hours of annealing, whereas the monolithic Nb and the Zr/Nb NMMs with $L$ = 10 nm and $L$ = 75 nm showed no signs of degradation even after 2 weeks of annealing. This behaviour was due to the much slower oxidation rate of Nb, which allowed the accommodation of the large residual stresses induced by the volume increase associated with the oxidation process by the plastic deformation on Nb [46]. The hardness of Zr/Nb for $L$ = 75 nm increased with annealing time from ≈ 7.7 GPa in the as-deposited condition to values in between those for monolithic Zr and Nb: The maximum hardness of ≈ 10.9 GPa was achieved after $t_a$ = 15 h, and it remained approximately constant afterwards. In the case of Zr/Nb for $L$ = 10 nm, the hardness increased from ≈ 7.2 GPa to ≈ 12.4 GPa after 2 hrs of annealing, and increased slightly up to ≈ 13 GPa after 336 hrs of annealing, even above the rule of mixtures between $ZrO_2$ and $Nb_2O_5$. As demonstrated before [20, 46], the large increase in the hardness after annealing arose from the oxidation of the Zr and Nb layers and the additional contribution of the residual stresses that develop as a result of the associated volumetric expansions. The much larger increase in hardness observed for $L$ = 10 nm in comparison to $L$ = 75 nm is associated to the development of damage in the form of voids and cracks within the $ZrO_2$ layers and along the interfaces between $ZrO_2$ and $Nb_2O_5$ (Fig. 8a) in the NMM with $L$ = 75 nm, as a result of the volumetric expansion associated with the progressive oxidation of the layers [22]. On the contrary, these voids/crack were not found in the NMM with $L$ = 10 nm (Fig. 8b), for which only defects formed during the deposition process were observed [18], demonstrating the superior damage tolerance associated with layer thickness reduction.

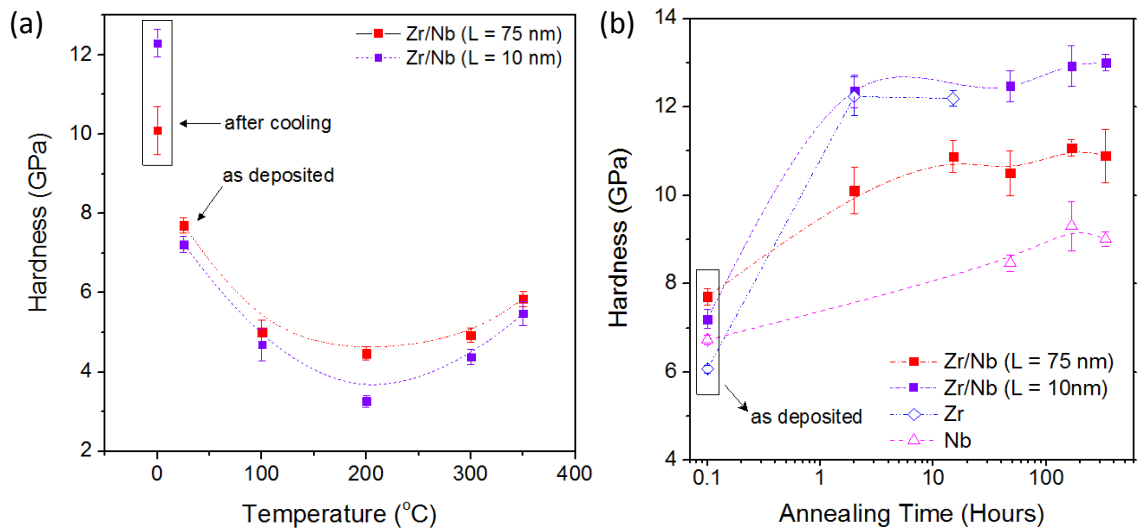

**Fig. 7.** (a) Indentation hardness as a function of temperature for Zr/Nb NMMs with periodicities $L$ = 10 and 75 nm. The "after cooling" values were measured at room temperature after cooling down from 350°C. (b) Indentation hardness as a function of annealing time for Zr/Nb NMMs with periodicities $L$ = 10 and 75 nm and for monolithic Zr and Nb films. The curves are guides to the eye.

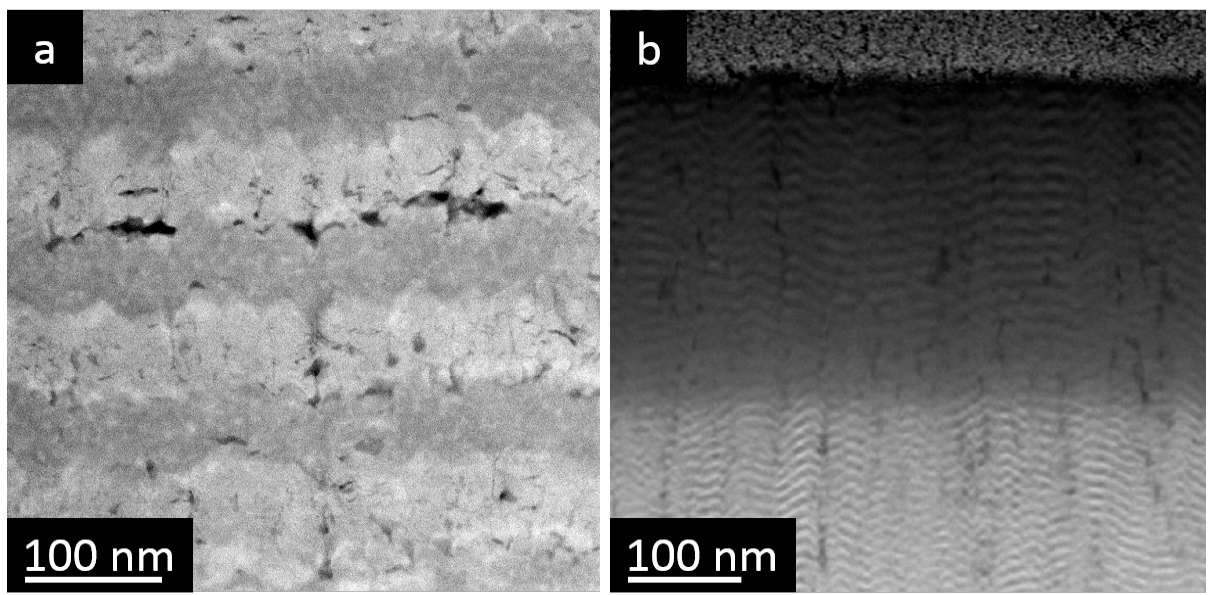

**Fig. 8.** HAADF-STEM images of Zr/Nb NMMs with $L$ of (a) 75 nm and (b) 10 nm annealed at 350°C for 168 hrs.

3.2.2. Micropillar compression tests

SEM images of the Zr/Nb micropillars, made of as-deposited and annealed (for 48 hrs) Zr/Nb multilayers, after compression are shown in Figs. 9a-h. Deformed Zr/Nb micropillars with $L$ = 75 nm in the as-deposited condition show evidence of initiation of plastic barrelling together with a few vertical cracks in the top half region of the pillar for strain levels below 15% (see Fig. 9a). After very large compressive strain (≈50%), the pillar collapses without any sign of shearing (Fig. 9b). There are some indications of extrusion observed in the side walls towards the bottom half of the pillar, while a large crack opens up in the top half of the pillar (see Fig. 9b). Plastic barrelling was also found in the case of compressed Zr/Nb micropillars with $L$ = 75 nm annealed for 48 hrs (Figs. 9c-d), although barrelling was localized in the middle of the pillar and became much more obvious as the applied strain increased. As-deposited and annealed micropillars with $L$ = 75 nm collapsed via crack opening along porous columnar grain boundaries [23].

Neither surface cracks nor significant barrelling was observed in the case of compressed Zr/Nb micropillars with $L$ = 10 nm in the as-deposited condition at low strains (Fig. 9e). The micropillars deformed homogeneously after deformation at very high strains (> 40 %), and plastic barrelling was observed. The pillars were bent sideways due to a slight miss-alignment of the punch during the compression test and cracks appeared near the pillar top in some cases (Fig. 9f). Finally, barrelling was not observed in the Zr/Nb micropillars with $L$ = 10 nm annealed for 48 hrs. These pillars appeared to deform mainly by shear (see Figs. 9g-h). Detailed structural analyses on compressed micropillars are presented in section 3.2.3.

Representative engineering stress-strain curves obtained from the micropillar compression tests are reported in Fig. 10. All curves showed an initial linear elastic response until the yield point was reached, followed by strain hardening. The apparent strain hardening response can be attributed to the increase in cross-sectional area of the micropillars during compression. The initial yield stress ($\sigma_{0.2\%}$) and the flow stress at $\varepsilon = 0.07$ ($\sigma_{7\%}$) obtained from the curves are listed in Table 1. The initial yield stress in the as-deposited condition was very similar for Zr/Nb NMMs with $L$ = 10 nm and 75 nm. It increased by ≈ 25% for $L$ = 10 nm and 75 nm after 48 hrs of annealing at 350ºC, whereas the flow stresses after the initial yield were larger for the annealed Zr/Nb pillars with $L$ = 10 nm.

The higher strength obtained for the $L$ = 10 nm nanolaminate is in agreement with the nanoindentation results and is also attributed to the presence of defects at the interfaces and inside the layers for $L$ = 75 nm. However, it is interesting to notice that the increase in flow stress with annealing for an equivalent strain of 7 % obtained from micropillar compression ($\approx$ 34% and $\approx$ 22 %) is not as large as the hardness increase measured by nanoindentation ($\approx$ 72% and $\approx$ 33 %) for $L$ = 10 nm and 75 nm, respectively. This is mainly due to the different role played by the residual stresses developed during oxidation of Zr and Nb layers in each test. While the residual stresses have to be overcome during nanoindentation, leading to large increase in hardness which is added to the contribution associated with the higher hardness of $ZrO_2$ and $Nb_2O_5$, they are largely released during FIB milling of the micropillar, and therefore do not contribute to the strength of the oxidized micropillars.

As a result, the effect of internal stresses in strengthening the Zr/Nb NMMs during nanoindentation is evident in the case of the Zr/Nb NMM with $L$ = 10 nm, where the hardness increase is almost twice the strength increase observed from micropillar compression tests. In the case of the Zr/Nb NMM for $L$ = 75 nm, the hardness increases after 48 hrs annealing is much smaller because of the development of damage at the interfaces after Nb is transformed into $Nb_2O_5$, as shown in Fig. 8a, which reduces the hardness and releases some residual stresses [22]. This damage was not found in the Zr/Nb NMM with $L$ = 10 nm (Fig. 8b), which exhibited a very large nanoindentation hardness increase after long term annealing (Fig. 7b). Therefore, the increase in strength observed in micropillar compression tests after annealing during 48 hrs at 350ºC is mainly a result of the higher mechanical properties of the oxide layers.

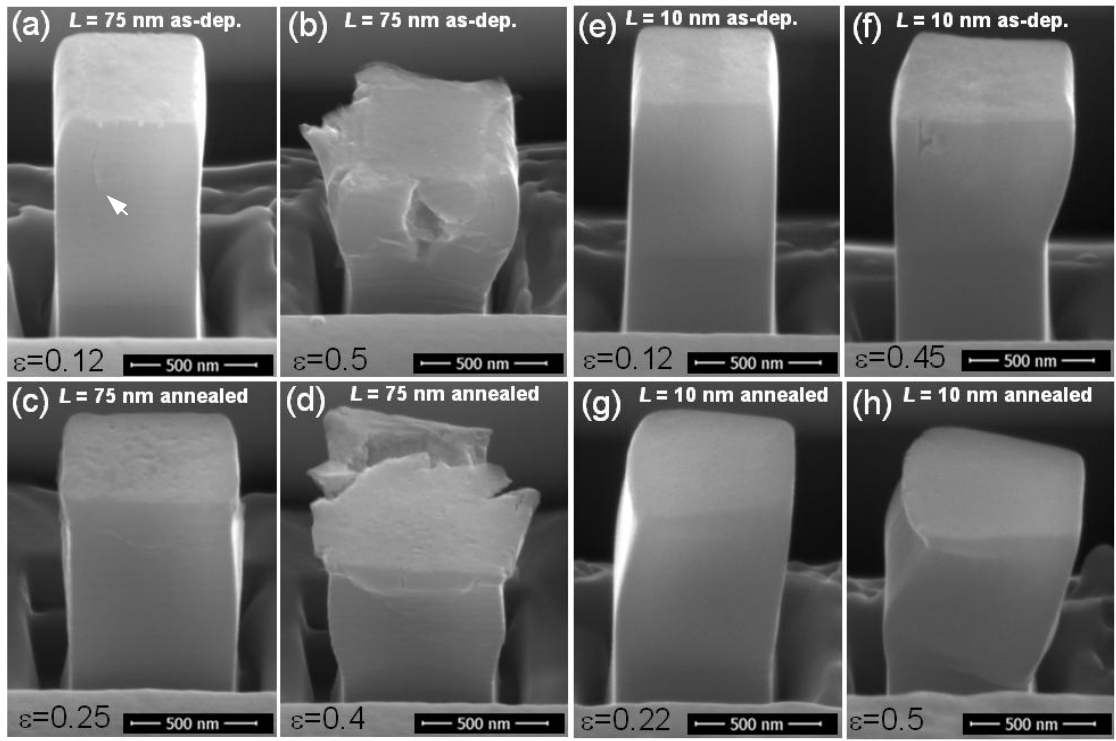

**Fig. 9.** SEM images of compressed Zr/Nb micropillars for different levels of strain: (a) and (b) Zr/Nb ($L = 75$ nm) as-deposited; (c) and (d) Zr/Nb ($L = 75$ nm) annealed for 48 hrs; (e) and (f) Zr/Nb ($L = 10$ nm) as-deposited; (g) and (h) Zr/Nb ($L = 10$ nm) annealed for 48 hrs.

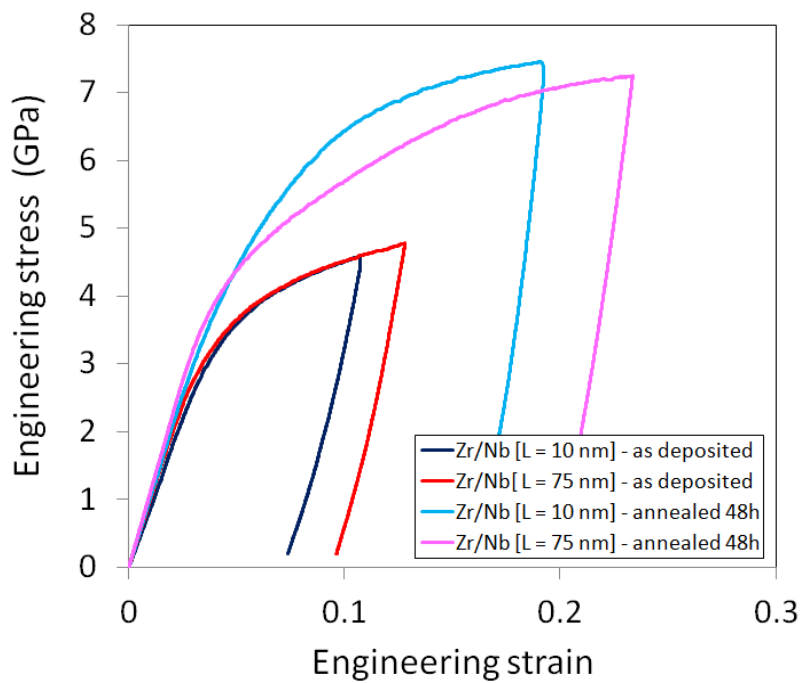

**Fig. 10.** Engineering stress-strain plots obtained from micropillar comparison tests on the Zr/Nb NMMs ($L = 10$ and $75$ nm) as-deposited and annealed for 48 hrs at 350ºC.

**Table 1.** Initial yield stress ($\sigma_{0.2\%}$) and flow stress at $\varepsilon = 0.07$ ($\sigma_{7\%}$) of the as-deposited and annealed (for 48 hrs) Zr/Nb NMMs ($L$ = 10 nm and 75 nm) from micropillar compression tests. $\sigma_{0.2\%}$ values obtained from finite element simulations of the micropillar compression tests are also included.

| Sample | $\sigma_{0.2\%}$ (GPa) | $\sigma_{7\%}$ (GPa) |
|---|---|---|
| Zr/Nb ($L$ = 10 nm) as-deposited | 2.9 ± 0.2 | 4.1 ± 0.2 |
| Zr/Nb ($L$ = 10 nm) annealed 48 hrs | 3.6 ± 0.2 | 5.4 ± 0.1 |
| Zr/Nb ($L$ = 75 nm) as-deposited | 2.8 ± 0.2 | 4.1 ± 0.2 |
| Zr/Nb ($L$ = 75 nm) annealed 48 hrs | 3.5 ± 0.2 | 5.0 ± 0.2 |

*3.2.3. Nanoscale plasticity of deformed Zr/Nb micropillars*

TEM samples of compressed micropillars were produced by FIB and investigated in detail by analytical electron microscopy to understand in much greater detail the deformation mechanism in the as-deposited and annealed Zr/Nb NMMs with different periodicities. The cross-sectional STEM image of the compressed (as-deposited) Zr/Nb NMM with $L$ = 75 nm is shown in Fig. 11a. This image revealed a symmetrical barrelling in the top region of the pillar (~ 16 layers), which was attributed to the formation and propagation of cracks along columnar grain boundaries, especially in the top region (highlighted by arrows in Fig. 11a). These cracks showed the reduced cohesive strength between columnar grains which facilitated the barrelling effect during compression. Although a small rotation of the layers was observed in the top-left corner of the pillar (Fig. 11b), it did not trigger the formation of shear bands. Beside crack formation, fracture of the layers was occasionally observed, especially for Nb layers located in the top region of the micropillar (highlighted in Fig. 11b) but no interfacial delamination occurred across the micropillar. Moreover, no layer mixture occurred after compression, as shown in Fig. 11c (oxygen map is reported for comparison with annealed samples).

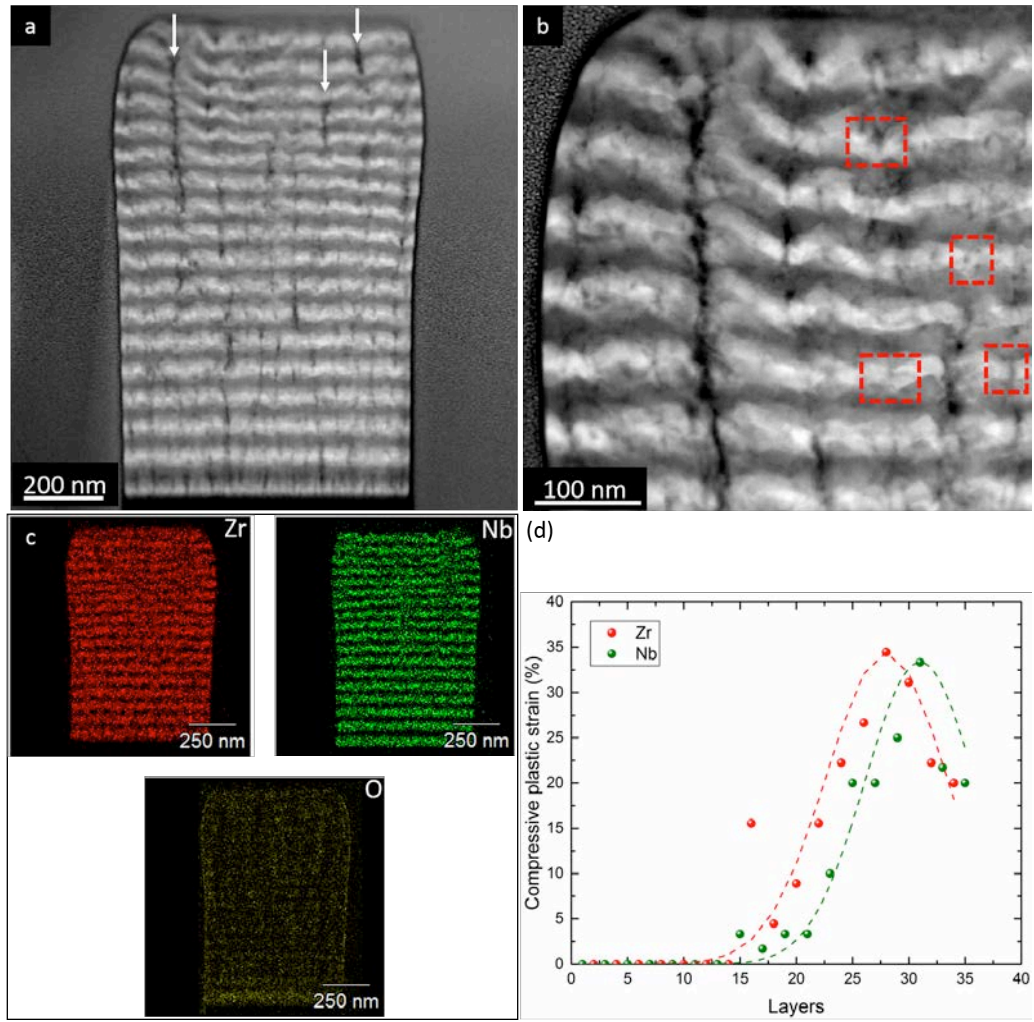

**Fig. 11.** Compressed micropillar of as-deposited Zr/Nb ($L$ = 75 nm): (a) cross-sectional HAADF-STEM image of the compressed micropillar (arrows point cracks along columnar grain boundaries); (b) HAADF-STEM image of the top-left region, where ruptures in Nb layers are highlighted; (c) EDX maps of the compressed micropillar showing the Zr, Nb and O content. (d) Compressive plastic deformation of the micropillar calculated from bottom (layer 0) to top (layer 36).

In order to understand the deformation mechanism in this multilayer, plastic deformation of the layers across the micropillar was evaluated. In particular, the individual thickness of each layer was measured along the loading axis over a region 100 nm wide to get a representative measurement. The compressive plastic strain was calculated from the thickness of the layers in the as-deposited condition ($t_{Nb}$ = 30 nm and $t_{Zr}$ = 45 nm), and reported as a function of the number of layers (Fig. 11d). Plastic deformation was concentrated on the layers in the upper half of the micropillar in both elements, and the distribution of plastic strain showed a similar trend, although Zr layers underwent a slightly higher plastic strain compared to Nb. However, the maximum strain (31 – 33%) was similar for both elements and was reached in the third

Nb layer and in the forth Zr layer from the top surface. These results are in agreement with the fact that monolithic Zr and Nb layers exhibit comparable yield stresses ($\sigma_{Nb}^{y} \approx$ 2.5 GPa and $\sigma_{Zr}^{y} \approx$ 2.2 GPa [46]) and co-deformation of both elements is expected during micropillar compression (Fig. 13b). Co-deformation of the constituent layers was reported for other multilayer systems (known as "load-bearing" effect [54]). Based on this model, Zr and Nb layers can be seen as two layers with similar mechanical properties and therefore, these considerations lead to assume similar deformation mechanisms in Zr and Nb layers.

A more complex scenario was found on compressed micropillars of Zr/Nb NMMs ($L$ = 75 nm) annealed for 48 hrs, which exhibit a double barrel shape, as shown in the HAADF-STEM image of Fig. 12a. In particular, three different regions can be distinguished: (i) the top region where heavy oxidation of the layers occurred, (ii) the middle region where uniform deformation took place, and (iii) the barrelled region at the bottom of the micropillar. The differences in the oxygen content of these regions are shown in Fig. 12c. Full oxidation on the Zr and Nb layers is observed in region (i) while only Zr is fully oxidized in region (ii) and the oxygen content was much lower in both Zr and Nb layers in region (iii). The plastic deformation in each layer was measured as in the previous case by comparison of the actual layer thickness with the thickness of the layers in the as-deposited or as-oxidized condition ($t_{ZrO2}$ = 54 nm and $t_{Nb2O5}$ = 47 nm) and it is depicted in Fig. 12d. Plastic deformation was accumulated in region (i) and (iii). In the top region (i), a uniform deformation of the oxidised layers was observed (with no preferential extrusion), although the deformation mechanisms were different: $Nb_2O_5$ layers showed plastic deformation while damage in the form of cracks and voids was observed in the $ZrO_2$ layers (see arrows Fig. 12b). These cracks were very likely present before the deformation of the micropillar and grew during deformation (see arrow Fig. 9a). Co-deformation of the Zr and Nb layers, leading to the barrelling of the micropillar in this region, was observed in region (iii). In the middle region (ii), uniform deformation of both layers was found but the plastic deformation was smaller than in region (iii), particularly in the Zr layers. This is very likely due to the oxidation of Zr to $ZrO_2$ in this region (fig. 12b) and also to the higher content of oxygen in Nb which could increase the flow stress. Therefore, based on the yield stresses of the different phases ($\sigma_{Nb}^{y} \approx$ 2.5 GPa, $\sigma_{Zr}^{y} \approx$ 2.2 GPa, $\sigma_{ZrO2}^{y} \approx$ 4.2 GPa and $\sigma_{Nb2O5}^{y} \approx$ 3.1 GPa [46]), it can be concluded that deformation was first localized in the barrelled region (iii), while deformation in the top region (i) took place only once a certain strain hardening was reached in the barrelled region.

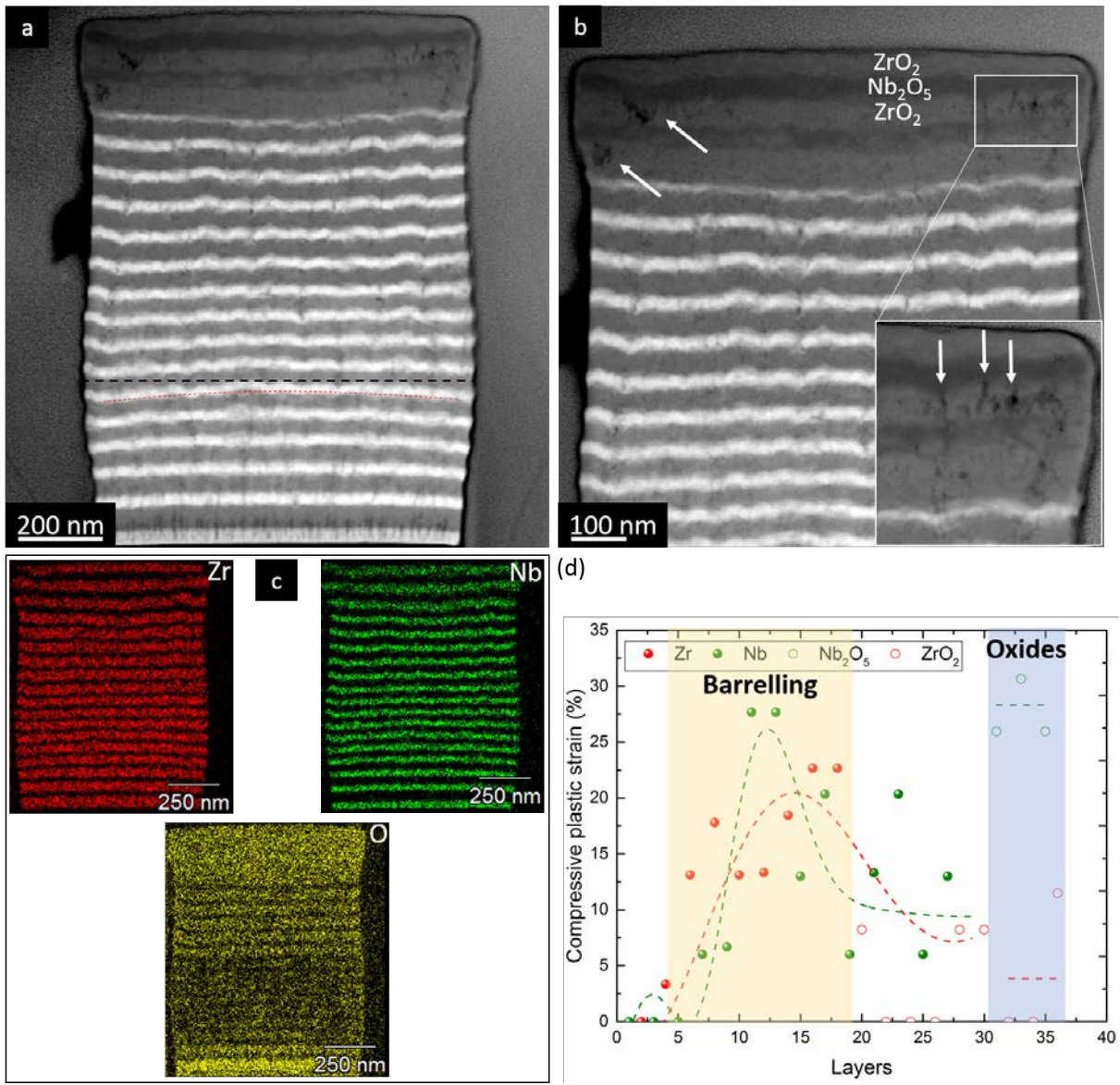

**Fig. 12.** Compressed micropillar of Zr/Nb NMMs ($L$ = 75 nm) annealed for 48 hrs: (a) cross-sectional HAADF-STEM image (dashed lines highlight bending of the layers); (b) HAADF-STEM image of the top and side (the arrow indicates the film growth direction) region of the compressed micropillar; (c) EDX maps of the compressed micropillar showing the Zr, Nb and O content. (d) Compressive plastic deformation of the micropillar calculated from bottom (layer 0) to top (layer 36).

Zr/Nb NMMs with $L$ = 10 nm have a complex structure which includes a pronounced columnar structure, porosity along grain boundaries and prominent layers waviness. A deformed micropillar for the Zr/Nb NMMs with $L$ = 10 nm in the as-deposited condition is shown in Fig. 13.

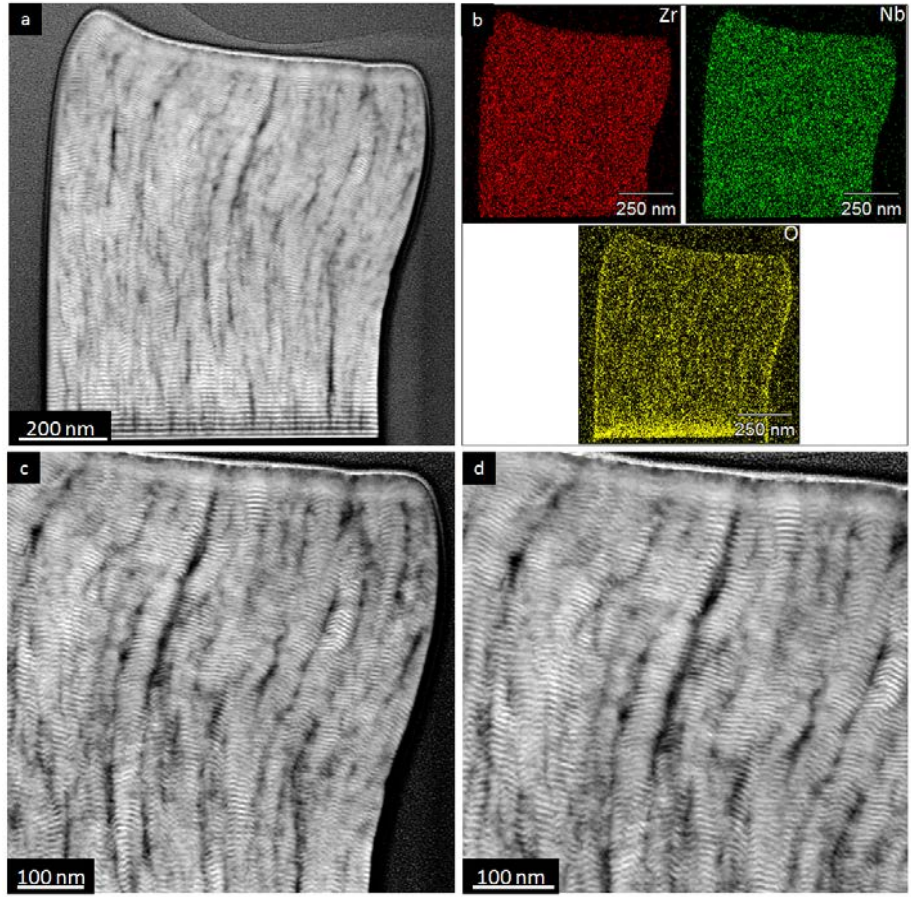

**Fig. 13.** Compressed micropillar of as-deposited Zr/Nb ($L$ = 10 nm). (a) Cross-sectional HAADF-STEM image of the compressed micropillar. (b) EDX maps of the micropillar showing the Zr, Nb and O content. (c) HAADF-STEM image of the top-right region of the micropillar. (d) HAADF-STEM image of the top-middle region of the micropillar.

The deformed pillar shows some inhomogeneous deformation near the top of the pillar, due to a slight misalignment of the flat punch. Micropillar compression caused plastic barrelling and preferential bending of the columns (Fig. 12c-d), while no disruption of the layered structure was observed (Fig. 12c). This indicates that voids along grain boundaries did not form following the compression test, as it was the case in the NMM with $L$ = 75 nm (Fig. 11). Plastic strain for Zr and Nb layers varied because of the variable individual layer thickness ($t_{Nb}$ = 2 – 5 nm and $t_{Zr}$ = 4 – 7 nm) caused by the pronounced waviness. For instance, layers as thin as 2 – 3 nm and 4 – 4.5 nm were measured for Nb and Zr, respectively, in the top-left region of the micropillar (Fig. 13c) where more severe deformation occurred. Plastic strains in the range 14 – 43% and 18 – 27% were computed for Nb and Zr layers, respectively, assuming an average thickness of 3.5 and 5.5 nm for Nb and Zr, respectively.

The deformed micropillar of the Zr/Nb NMM ($L$ = 10 nm) annealed for 48 hrs is shown in Fig. 14a. Deformation was mostly localised in the top-left and bottom-right regions of the micropillar (Fig. 14c and 14d), where shear bands are shown to form at an angle of about 35 – 37° away from the loading direction. The shear band formed in the top-left region (inset in Fig. 14c) did not extend across the boundary (about 140 nm below the top surface). The shear band formed on the bottom-right region (Fig. 14d) extended for about 250 nm above the substrate. By considering the higher yield stress exhibited by oxides (mostly located in the top region according to the EDX map in Fig. 14b-c) and the extension of the shear bands, the one located on the bottom-right formed first and accommodated most of the plastic deformation.

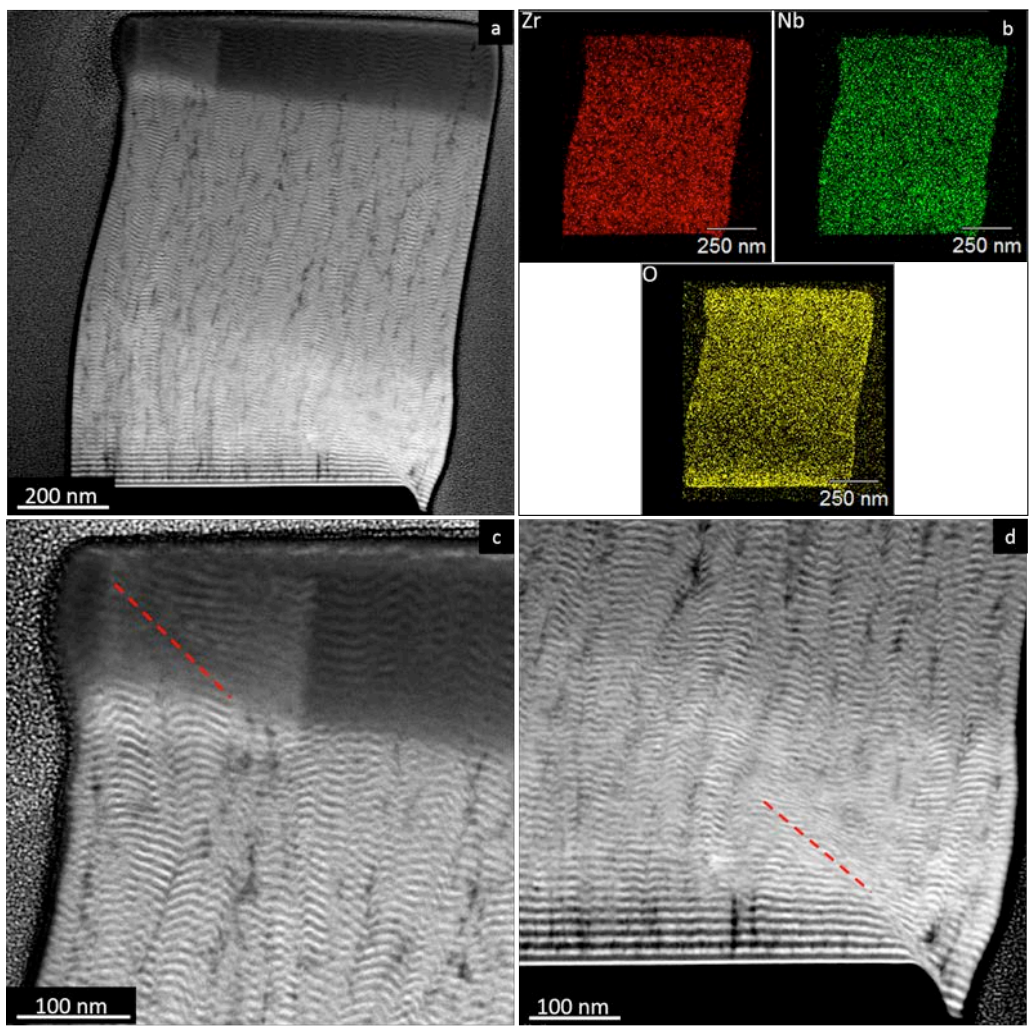

**Fig. 14.** Compressed micropillar of Zr/Nb (L = 10 nm) annealed for 48 hrs. (a) Cross-sectional HAADF-STEM image. (b) EDX maps of the whole micropillar showing Zr, Nb and O content. (c) HAADF-STEM image of the shear band formed on the top-left region of the pillar. (d) HAADF-STEM image of the shear band formed on the bottom-right region of the pillar.

## 4. Conclusions

Contrary to most nanoscale multilayers that show a limited thermal stability and degradation under oxidizing environments, annealing in air at 350ºC of magnetron-sputtered Zr/Nb nanoscale multilayers leads to the formation of strong and stable oxide/metal nanocomposite coatings, with a well-defined layered structure and sharp interfaces. Moreover, a reduction in bilayer thickness from 75 nm to 10 nm resulted in superior strength and damage tolerance of the oxidized coatings. The following conclusions are drawn regarding the oxidation process and the corresponding mechanical properties:

- The hardness and strength of the as-deposited NMMs was independent of the layer thickness. Additionally, the plastic strain measured on compressed micropillars (in as-deposited condition) revealed co-deformation of both Zr and Nb layers regardless of the bilayer thickness.
- Annealing of Zr/Nb in air led to transformation of the Zr layers into $ZrO_2$ after a few hours at 350 ºC, while the Nb layers oxidised at a much slower rate. The sequential oxidation of Zr and Nb layers was found to be key for the oxidation to take place without rupture of the multi-layered structure and without coating spallation in all cases, because the volumetric expansion associated with the oxidation of the Zr layers was partially relieved by the metallic Nb layers.
- A bilayer thickness reduction from $L = 75$ nm to $L = 10$ nm results in a superior stability of the layered structure during the oxidation process and a larger final strength. This is because the volumetric expansion associated with the progressive oxidation of the Nb layers led to the formation of cracks at the interfaces and within the $ZrO_2$ layers for the NMMs with $L = 75$ nm but not in the case of $L = 10$ nm.
- The enhanced structural integrity observed in NMMs with $L = 10$ nm led to a more pronounced hardness increase (≈72 %) compared to $L = 75$ nm (≈ 33%). The main sources of the increase in hardness were the compressive residual stresses generated upon annealing because of the volumetric expansion and the higher hardness of the $ZrO_2$ and $Nb_2O_5$ oxides. Relaxation of the residual stresses in the NM with $L = 75$ nm due to damage explained the lower hardness, as compared with $L = 10$ nm.
- The strength increase measured in the micropillar compression tests after annealing was not as large as the hardness increase (≈ 34% for $L = 10$ nm and ≈ 22 % for $L = 75$ nm) because the residual stresses were released during FIB milling of the micropillars. Therefore, the pillar strengthening was mainly a result of the higher mechanical properties of the oxidised Zr and Nb layers, without contributions from residual stresses.

**Acknowledgements**

This investigation was supported by the European Research Council (ERC) under the European Union's Horizon 2020 research and innovation programme (Advanced Grant VIRMETAL, grant agreement No. 669141, from the Madrid region under programme S2013/MIT-2775 (DIMMAT-CM), and by Czech Science Foundation, grant no. 17-17921S. M. C. and T. P. acknowledge EPSRC program grant EP/K040375/1 'South of England Analytical Electron Microscope'.